\documentclass[12pt,showpacs]{iopart}

\usepackage{graphics,graphicx}
  \DeclareGraphicsExtensions{.eps}   
  \graphicspath{{figures/}}

\usepackage{iopams}
\usepackage[percent]{overpic}

\newcommand{\TG}{\tilde{\Gamma}}
\newcommand{\psimm}{\psi^{-2}}
\newcommand{\er}{\eta_s(\vec{r})}
\newcommand{\etas}{\eta_s}
\newcommand{\gt}{\tilde{\gamma}}
\newcommand{\order}[1]{$ #1 ^{\mbox{\small th}}$}

\begin{document}

\title{
Toward a dynamical shift condition for unequal mass black hole
binary simulations
}

\author{Doreen M\"{u}ller, Bernd Br\"{u}gmann}

\address{Theoretical Physics Institute, 
University of Jena, 07743 Jena, Germany}

\date{\today}

\begin{abstract}
Moving puncture simulations of black hole binaries rely on a specific
gauge choice that leads to approximately stationary coordinates near
each black hole. Part of the shift condition is a damping parameter,
which has to be properly chosen for stable evolutions. However, a
constant damping parameter does not account for the difference in mass
in unequal mass binaries. We introduce a position dependent shift
damping that addresses this problem. Although the coordinates change,
the changes in the extracted gravitational waves are small. 
\end{abstract}
\pacs{
04.25.D-, 
04.25.dg,
04.30.Db 
}

\section{Introduction}
\label{sec:Introduction}

Early numerical relativity simulations using a 3+1 split of the
Einstein equations suffered from so-called slice stretching, an effect
which occurs when using singularity avoiding slicing together with a
vanishing shift. The slices become highly distorted when time marches
on in the outer regions of the grid but slows down in the vicinity of
the black hole. It became clear that a non-vanishing, outwards
pointing shift vector would be required in order to redistribute grid
points and also to prevent grid points from falling into the black
hole.  Inspired by Balakrishna {\em et al.}~\cite{BalDauSei96},
Alcubierre {\em et al.}~\cite{AlcBru00,AlcBruPol01} combined the 1+log
slicing condition with a dynamical shift condition called gamma-driver.
These gauge conditions successfully
prevented slice stretching in black hole simulations using
excision. It turned out that such gauge conditions could be used also
for fixed punctures with slight modifications to keep the puncture
from evolving~\cite{AlcBruDie02}. The fixed-puncture modification was
removed in~\cite{CamLouMar05,BakCenCho05} when the moving puncture
method was introduced. 1+log slicing with gamma-driver shift succeeds
in moving the puncture freely through the grid while simultaneously
avoiding slice-stretching.  The basic reason for the success of this
gauge condition is that when the slices start to stretch, the shift
vector counteracts by pulling out grid points from the region near the
black hole.

In this paper we focus on the dissipation or damping parameter in the
gamma-driver shift condition, which plays an important role in the
success of this gauge. In order to reduce oscillations in the shift
vector, the authors of \cite{AlcBruPol01} noticed the necessity of a
damping term in the
shift condition. Adjusting the strength of the damping via a damping
parameter was found to allow freezing of the evolution at late times
in \cite{AlcBruDie02} and to avoid drifts in metric variables
in~\cite{BruGonHan06}. Additionally, the value of the damping
coefficient was found to affect the coordinate location of the
apparent horizon and therefore the resolution of the black hole on the
numerical grid ~\cite{BruGonHan06,HerShoLag06}. The right choice of
the damping value is therefore important if one wants to resolve the
black hole properly while still driving the coordinates to a frame
where they are stationary when the physical situation is stationary and hence
obtain a stable evolution.

The specific value of the damping parameter has to be adapted to the
black hole mass in order to obtain long term stable evolutions.  If
the damping parameter is either too small or too large, there are
unwanted oscillations or a coordinate instability, respectively. In
binary simulations, a typical choice is a constant value of roughly
$2/M$, where $M$ is the total mass of the system.
However, using a constant damping parameter for black hole binaries with
unequal masses leads to a fundamental problem.  With a constant
damping parameter, the effective damping near each black hole is
asymmetric for unequal black hole masses since the damping parameter
has dimensions $1/M$. For large mass ratios, this asymmetry in the
grid can be so large that simulations fail because the damping may
become too large for one of the black holes. This is one of the reasons why
the highest mass ratio that has been successfully simulated up to now
is $10:1$~\cite{GonSpeBru08}.

Advantageous would be a position-dependent damping parameter that
adapts to the local mass, in particular such that in the vicinity of
the \order{i} puncture with mass $M_i$ its value approaches $1/M_i$.  It
was noticed before \cite{AlcBruDie02,GonSpeBru08} that a damping
coefficient adapted to the various parameters of the simulation would
be beneficial. In \cite{AlcBruDie04} a position-dependent formula was
introduced for head-on collisions of black holes, which to our
knowledge was only used in one other publication \cite{ZloBakCam05},
prior to the moving puncture framework.  In this paper, we take first
steps towards a position-dependent damping parameter for moving
punctures. As a consequence, the local coordinates change compared to
standard simulations, but this does not significantly affect gauge
invariant quantities like the extracted waves as we discuss below.

\section{Dynamical damping in the shift equation}

  \subsection{Numerical setup}

  We focus on the gauge condition used in the 3+1 splitting of
  the Einstein equations, in particular on the condition for the shift vector.

  The slices are determined by the 1+log slicing condition \cite{BonMas94} for
  the lapse function $\alpha$,
  \begin{equation}
    \partial_0 \alpha = -2\alpha K,
    \label{eq:onepluslog}
  \end{equation}
  where $K$ is the trace of the extrinsic curvature.
  The coordinates of a given slice are governed by the gamma-driver shift
  condition introduced in \cite{AlcBruDie02} as
  \begin{equation}
    \partial_{0}^2 \beta^i = \frac{3}{4} \partial_0 \TG^i - \etas \partial_0
    \beta^i,
    \label{eq:gammadriver}
  \end{equation}
  where  $\TG^i$ are the contracted Christoffel symbols of the conformal metric
  $\gt_{ij}$, $\beta^i$ is the shift vector and $\etas$ is the damping
  coefficient  we will discuss in this publication.
  In Eqs.~(\ref{eq:onepluslog}) and (\ref{eq:gammadriver}), $\partial_0$ is
defined as
  $  \partial_0 = \partial_t - \beta^i\partial_i $
  as suggested by \cite{BeySar04,GunGar06,MetBakKop06}. 

  Examining the physical dimensions, we see that $[\beta^i] = 1$ and
$[\partial_0] = 1/M$, where $M$ is the mass (e.g.\ the total mass of the
spacetime under consideration).
  For this reason,  the second term on the right hand side of equation
  (\ref{eq:gammadriver}) requires the damping parameter to carry units,
  \begin{equation}
    [\etas] = \frac{1}{M}. 
    \label{eq:etaunits}
  \end{equation}

In simulations of a single Schwarzschild puncture of mass $M_1$, we
typically choose a damping parameter of $\eta_s \approx 1/M_1$ for
obtaining enough damping in the shift without producing
instabilities. In numerical experiments for a Schwarzschild puncture
(to be discussed elsewhere), we find that $0 \leq \eta_s \lessapprox
3.5/M_1$ is necessary for a stable and convergent numerical
evolution. Some minimal amount of damping is important to suppress noise
in the gauge when a puncture is moving. On the other hand, if $\eta_s$
is too large then there are gauge instabilities, leading to a loss of
convergence and to instability of the entire numerical evolution.
Furthermore, early simulations for fixed punctures also found that
$\eta_s$ should take values around $1/M$, where $M$ is the total mass, 
to avoid long-term coordinate drifts at the outer boundary \cite{AlcBruDie02}. 

In simulations of black hole binaries with total mass $M=M_1+M_2$, we
usually set $\etas = 2/M$ which has been found to work well in equal
mass binaries simulations. (For equal masses, $M_1=M_2=M/2$, so near
one of the punctures the value of $\etas$ discussed above for
Schwarzschild becomes $\etas=1/M_1=2/M$.)
For unequal mass binaries, the different black holes tolerate
different ranges of $\eta_s$ according to the above statement about
single black holes. Ideally, $\eta_s$ should be $\approx 1/M_i$, which
cannot be accomplished simultaneously for unequal masses using a
constant value of $\eta_s = 2/M$. In fact, for the mass ratio 1:10 in
\cite{GonSpeBru08}, the choice $\eta_s = 2/M$ failed, but a smaller
value for $\eta_s$ was chosen such that
$\eta_s\lessapprox3.5/M_i$ for both $i=1$ and $i=2$.

To overcome the conflicts between punctures with different masses in
evolutions of two or more black holes, we suggest to construct a
non-constant, position-dependent damping parameter which knows about
the position and mass of each puncture and takes a suitable value at
every grid point.
\subsection{Using $\psimm$ to determine the position of the punctures}
We thus desire a definition of $\eta_s$ which respects the unit
requirements found in Eq.~(\ref{eq:etaunits}) and which asymptotes to
specifiable values at the location of the punctures and at infinity.
Typical values are $\eta_s = 1/M_i$ at the $i^{\mbox{th}}$ black hole
and $\eta_s = 2/M$ at large distances.
This can be achieved by determining $\etas$ through a position dependent
function defined on the whole grid instead of using a constant as
before. We desire a smooth $\eta_s$ which avoids modes which travel at
superluminal speeds. Since we use the
Baumgarte--Shapiro--Shibata--Nakamura (BSSN) system of Einstein's
equations \cite{ShiNak95,BauSha98}, we want the form of $\etas$ to
depend only on the BSSN variables
in a way that does not change the principal part of the differential operators.

In this paper, we choose to use the conformal factor $\psi$, which contains
information about the locations and masses of the punctures.
  The formula we will use for determining the damping
  coefficient $\er$ is
  \begin{equation}
    \er = \hat{R}_0 \frac{\sqrt{\gt^{ij}\partial_i\psimm\partial_j\psimm}} 
    {(1 - \psimm)^2},
    \label{eq:etapsimm}
  \end{equation}
with $\gt^{ij}$ the inverse of the conformal 3--metric and $\hat{R}_0$ a
dimensionless constant. While $\psi$, $\gt^{ij}$, and
$\hat{R}_0$ are dimensionless, the partial derivative introduces
the appropriate dependence on the mass since $[\partial_i]=1/M$ and hence
$[\er] = 1/M$.

For a single Schwarzschild puncture of mass $M$ located at $r=0$ the
behavior of Eq.~(\ref{eq:etapsimm}) near the puncture and near
infinity is as follows.
According to \cite{HanHusPol06}, for small radii $r$ (near the
puncture) the conformal factor asymptotically equals

    \begin{equation}
       \psi^{-2} \simeq p_1 r
       \label{psim2smallr}
    \end{equation}
for a known constant $p_1$. The next to leading order behavior is less 
simple~\cite{Bru09}. 
The point $r=0$ corresponds to a sphere with finite areal radius $R_0$,
    \begin{equation}
      R_0 = \lim\limits_{r\rightarrow0}\psi^2 r = \frac{1}{p_1} = \hat{R}_0 M.
    \end{equation}
Numerically, $\hat{R}_0\approx1.31$. 
    The inverse of the conformal metric behaves like
    \begin{equation}
      \gt^{ij} \simeq \delta^{ij}. 
    \end{equation}
    Therefore, we find for small $r$
    \begin{equation}
      \sqrt{\gt^{ij}\partial_i \psi^{-2} \partial_j \psi^{-2}} 
      \simeq p_1 = \frac{1}{\hat{R}_0 M}
      \label{eq:eta_small_r_num}
    \end{equation}
    and
    \begin{equation}
      (1 - \psimm)^2 \simeq (1 - p_1 r)^2 \simeq 1
      \label{eq:eta_small_r_denom}
    \end{equation}
    when keeping only leading order terms in $r$.
    Equations (\ref{eq:eta_small_r_num}) and (\ref{eq:eta_small_r_denom})
    combine according to (\ref{eq:etapsimm}) to give
    \begin{equation}
      \etas(r=0) = 1/M.
    \end{equation}
For large $r$ we can expand the conformal factor in powers of $1/r$,
\begin{equation}
\psimm \simeq \left( 1 + \frac{M}{2 r} \right)^{-2}
       \simeq 1 - \frac{M}{r},
\end{equation}
resulting in
    \begin{equation}
\sqrt{\gt^{ij}\partial_i \psi^{-2} \partial_j \psi^{-2}} 
= \frac{M}{r^2}.
       \label{eq:abspsim2_larg_r}
    \end{equation}
and
    \begin{equation}
      \etas(r \rightarrow \infty) \simeq \hat{R}_0 \frac{M/r^2}{(M/r)^2} =
      \frac{\hat{R}_0}{M}.
    \end{equation}
In summary, Eq.~(\ref{eq:etapsimm}) leads to 
    \begin{equation}
      \etas(r) \rightarrow \left\{
                              \begin{array}{*{2}{l}}
                                1/M, & r \rightarrow 0 \\
                                \hat{R}_0/M, & r \rightarrow \infty \\
                              \end{array}
                          \right.
      \label{eq:eta_limits}
    \end{equation}
for a single puncture at $r = 0$.  Note that using
Eq.~(\ref{eq:etapsimm}) in Eq.~(\ref{eq:gammadriver}) does not affect
the principal part of (\ref{eq:gammadriver}). Therefore, the system
remains strongly hyperbolic, same as for $\etas=\mbox{const.}$
according to \cite{GunGar06,BeySar04}.

\section{\label{results}Results}

  Our Eq.~(\ref{eq:etapsimm}) analytically gives the desired $1/M$
  behavior near the puncture and near infinity for a single,
  non--spinning and non--moving puncture. Now it remains to be tested
  whether these properties persist in actual numerical simulations,
  especially for unequal mass binaries.

  Simulations are performed with the BAM code described
  in \cite{BruGonHan06,HusGonHan07}.
  The code uses the BSSN formulation of Einstein's
  equations and employs the moving puncture framework
  \cite{CamLouMar05,BakCenCho05}.
  Spatial derivatives are \order{6} order accurate and time integration is
  performed using the \order{4} order Runge--Kutta scheme.
  The numerical grid is composed of nested boxes with increasing resolution,
  where the boxes of highest resolution are centered around the black holes. 
  These boxes are advanced in time with Berger--Oligar time stepping
  \cite{BerOli84}.
  We are using puncture initial data with Bowen--York extrinsic curvature and
  solve the Hamiltonian constraint using a pseudospectral collocation method
  described in \cite{AnsBruTic04}. The momentum parameter in the Bowen--York
  extrinsic curvature is chosen such that we obtain quasi--circular
  orbits in our binary simulations using the method of \cite{WalBruMue09}.

  For binary simulations with unequal masses, we will use the mass
  ratio $q=M_2/M_1$ to denote the runs, $M_i$ being the bare mass of
  the \order{i} puncture.  The physical masses of the punctures (obtained after
  solving the constraints) differ by less than $10\%$
  from the bare masses for the orbits considered
  here, so the $\etas$ values derived for a single puncture should
  remain valid.
  When comparing simulations run with $\etas=2/M$ and $\er$ following
  Eq.~(\ref{eq:etapsimm}) we will refer to them as ``standard'' and ``new'' or
  ``dynamical'' gauge, respectively, throughout this paper.

  \subsection{\label{results:schwarzschild}Single Schwarzschild Black Hole}

\begin{figure}[tb]
    \centering
    \includegraphics[width=0.45\textwidth]{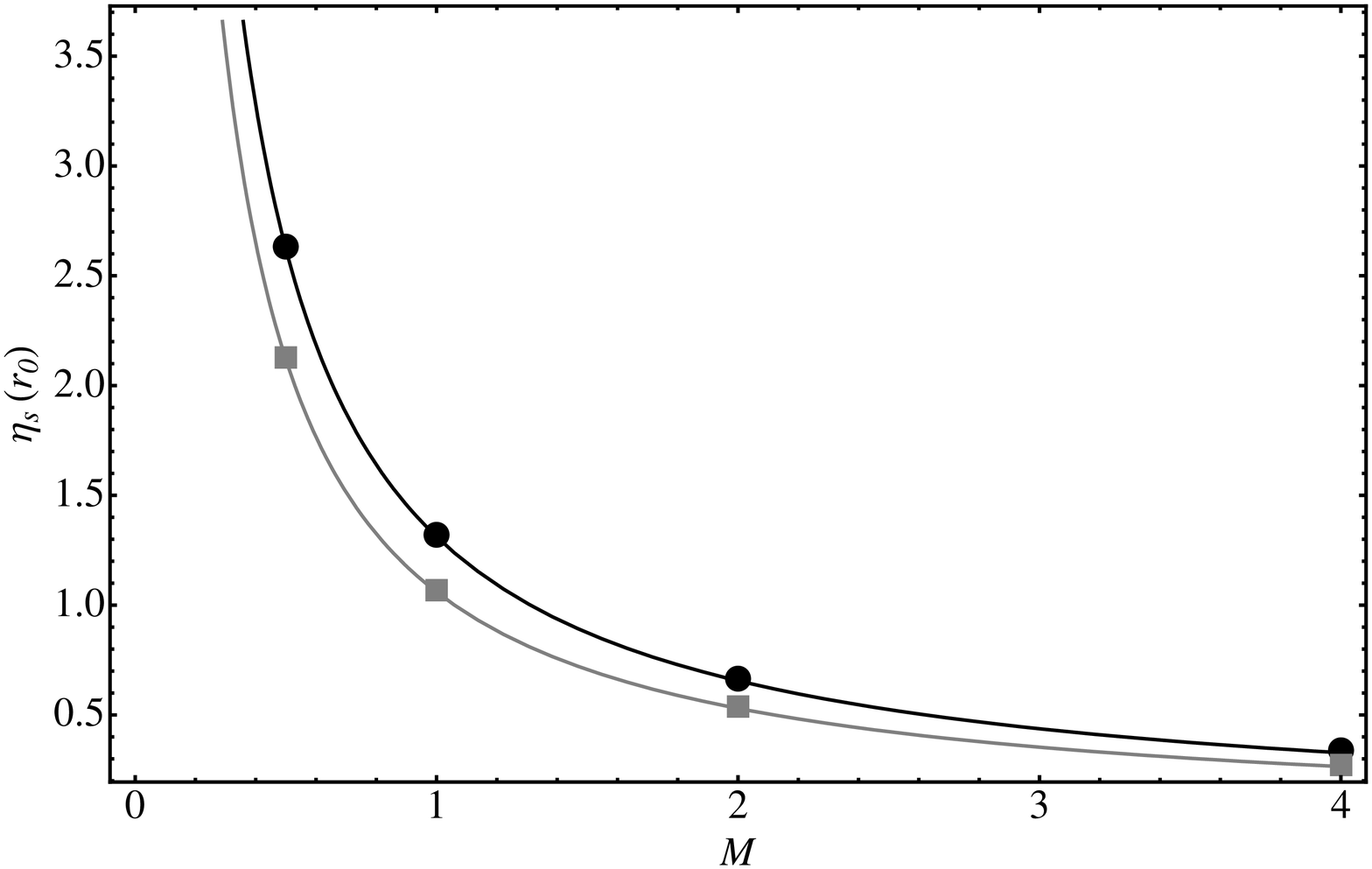}
    \put(-75,75){\includegraphics[width=0.15\textwidth]{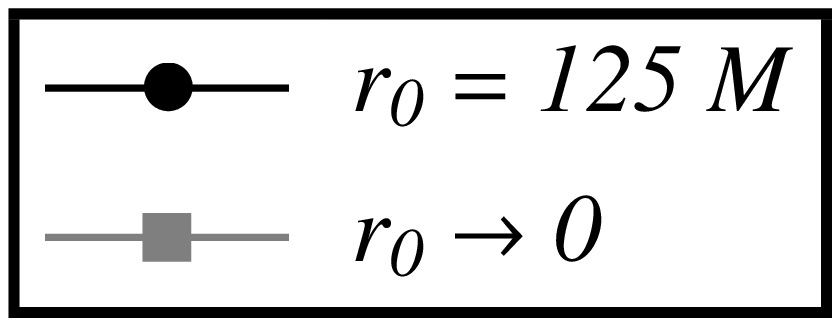}}
    \caption{
Numerical test of the analytical limits (\ref{eq:eta_limits}) of $\er$ using
single, non--spinning punctures with different masses.
Shown are the values of $\etas(r)$ near the puncture (gray squares) and at the
outer boundary (black dots). Note that here $M$ is identified with a dimensionless number, so $\etas(r)$ is dimensionless as well.
The fits to the data points are consistent with the analytic
prediction. Numerically, $\etas(M)=1.05/M$ (gray line) near the
puncture and $\etas(M)=1.311/M$ (black line) at the outer boundary.}
    \label{fig:eta_over_M}
  \end{figure}

  In order to test the $1/M$--behavior of (\ref{eq:etapsimm}) near the
  puncture and infinity, we first performed a series of evolutions for
  a time of $100\,M$ of a single, non--spinning puncture
  while varying its mass. We then measured the value of $\er$ near the
  puncture and at the outer boundary of the grid and compared these
  values to the limits (\ref{eq:eta_limits}).  The data points in
  Fig. \ref{fig:eta_over_M} correspond to these measurements while the
  lines are fits to the numerical data.  The values of $\er$ near the
  puncture as a function of total mass $M$ are fitted to
  $\etas(M)=1.05/M$ which agrees well with the analytical limit
  $r\rightarrow0$ of (\ref{eq:eta_limits}). Fitting to $\etas$ measured
  near the physical boundary of the grid reveals $\etas(M)=1.311/M$ and
  therefore fulfills the limit $r\rightarrow\infty$ of
  (\ref{eq:eta_limits}) even though the outer boundary is situated
  only at $130\,M$.

\begin{figure}[tb]
    \centering
    \includegraphics[width=0.45\textwidth]{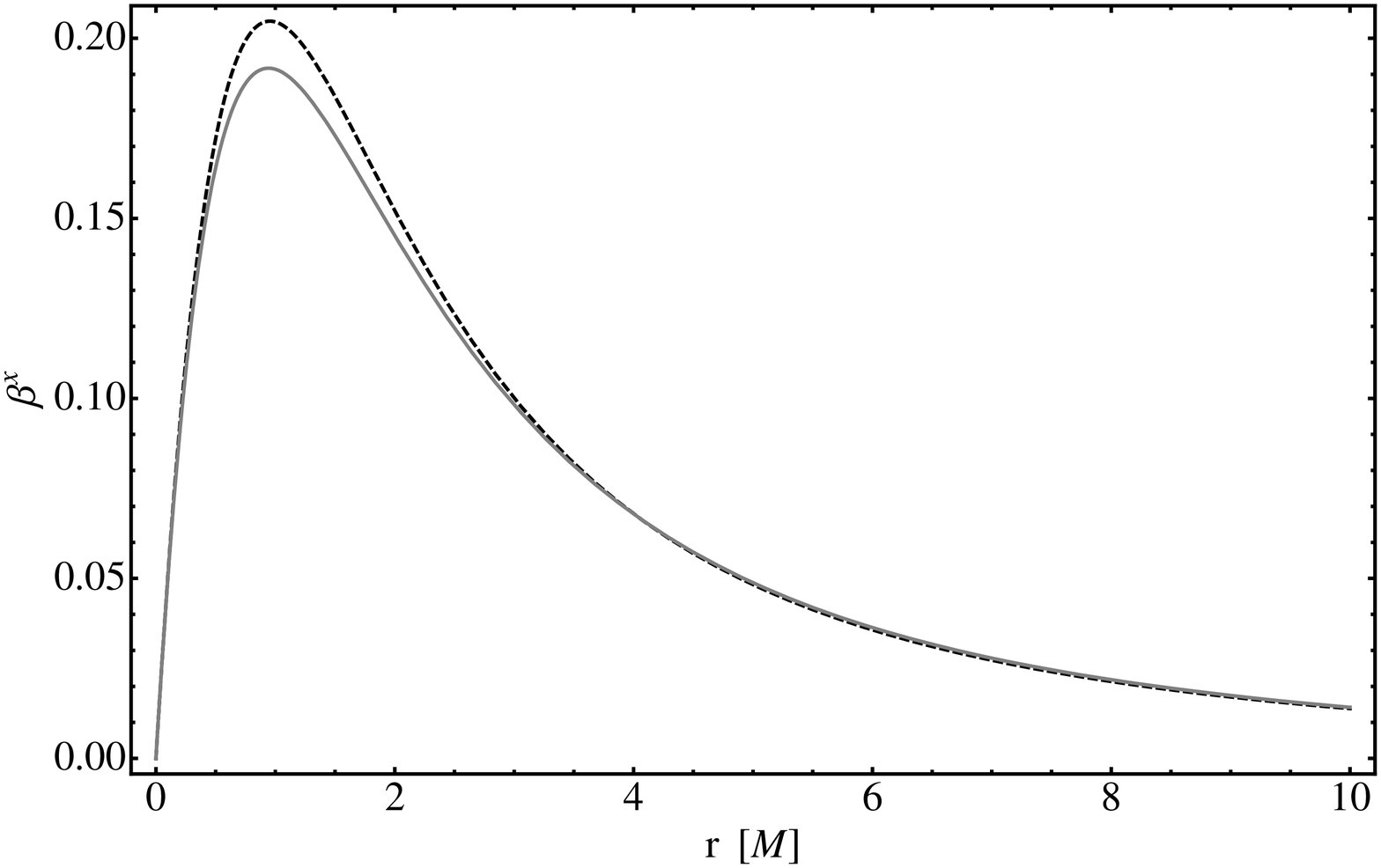}
    \put(-55,95){\includegraphics[width=.1\textwidth]{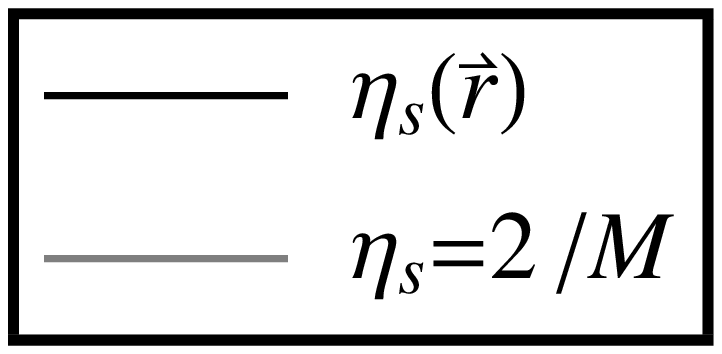}}
    \includegraphics[width=0.45\textwidth]{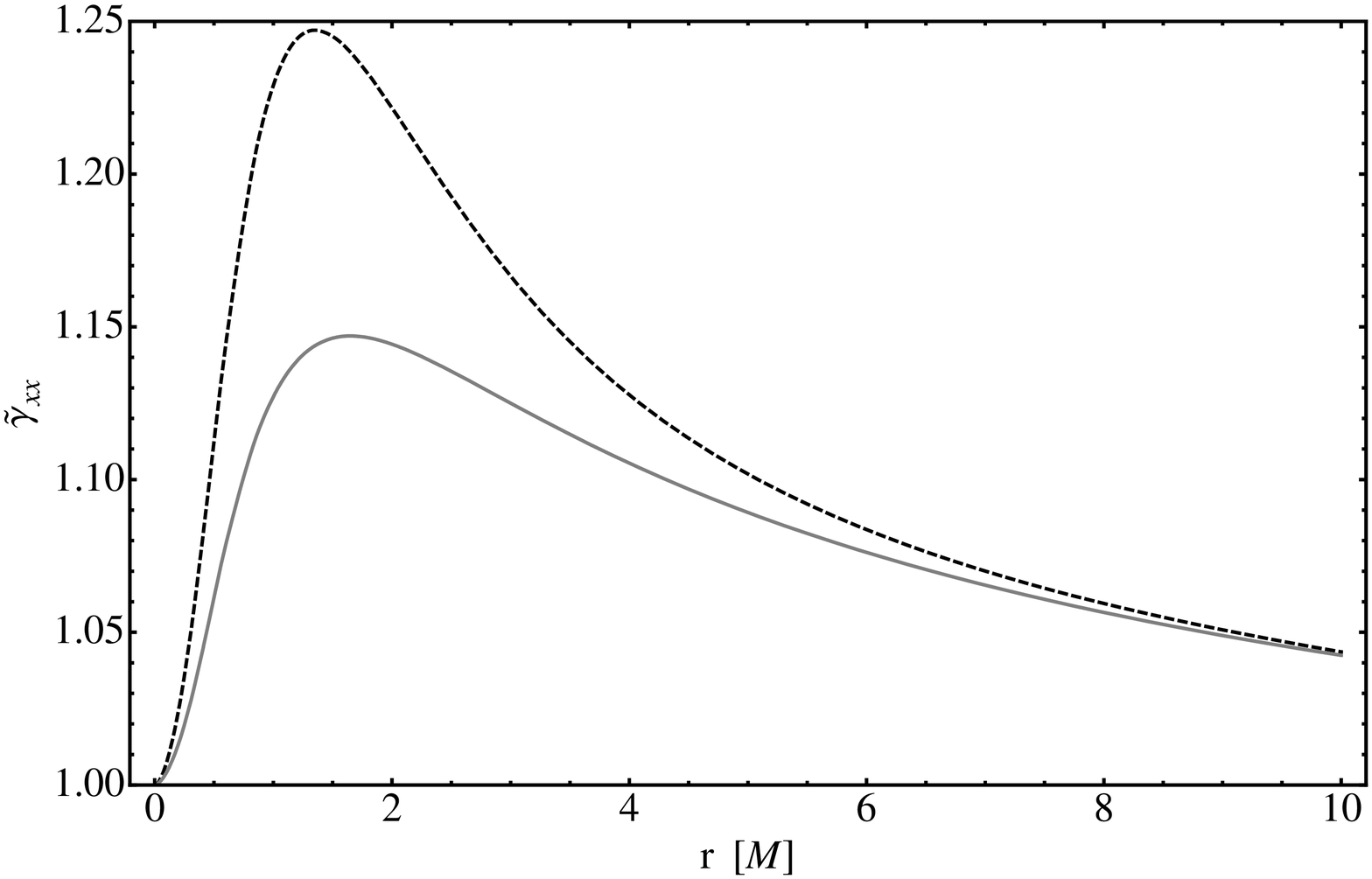}
    \put(-55,95){\includegraphics[width=.1\textwidth]{fig2_eta_legend2}}
    \caption{
    $x$-component of the shift vector (upper panel) and $xx$-component
    of the conformal 3-metric (lower panel) of a single, non-spinning puncture
    at time $t=100\,M$, where the simulations have reached a stationary state.
    The dashed black curves use dynamical damping, Eq.~(\ref{eq:etapsimm}), the
    gray ones use $\etas=2.0/M$ in the shift condition
    Eq.~(\ref{eq:gammadriver}).}
    \label{fig:betaxoverr}
  \end{figure}

  Using a modified shift condition, the shift itself will, of course, change. We
  compare the $x$-component of the shift vector for using $\etas=2.0/M$ and
  $\er$ in Fig.~\ref{fig:betaxoverr}. 
  A change in the shift implies a change of the coordinates and therefore,
  coordinate dependent quantities will change, too. As an example, the
  $xx$-component of the conformal 3-metric,
  $\tilde{\gamma}_{xx}$, is compared for $\etas=2/M$ and
  $\er$ in the lower panel of Fig.~\ref{fig:betaxoverr}. 
  The comparisons are made at time $t=100\,M$, when the simulations have
  reached a stationary state.

\begin{figure}
   \centering
    \includegraphics[width=0.45\textwidth]{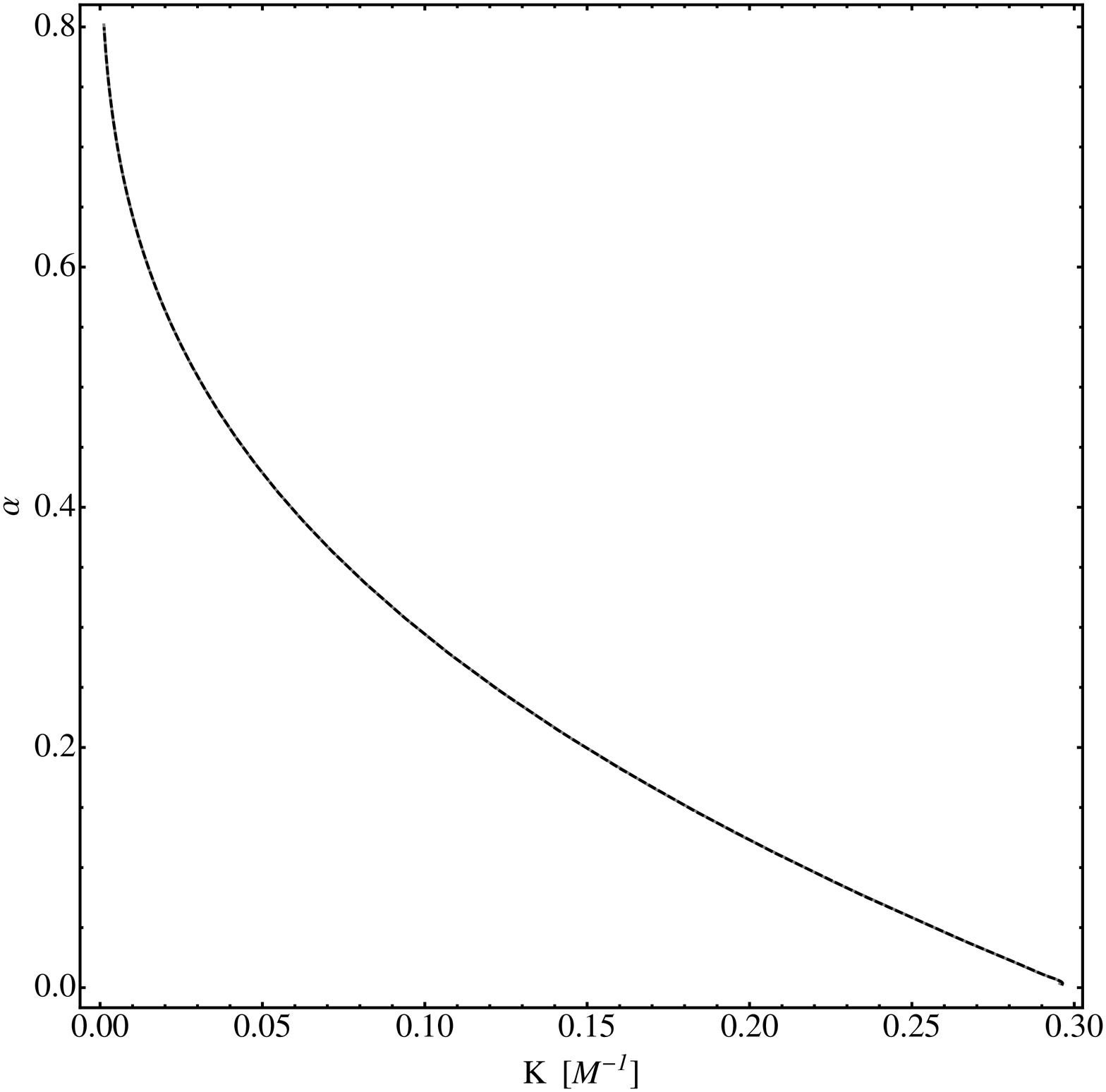}
    \put(-100,135){\includegraphics[width=.2\textwidth]{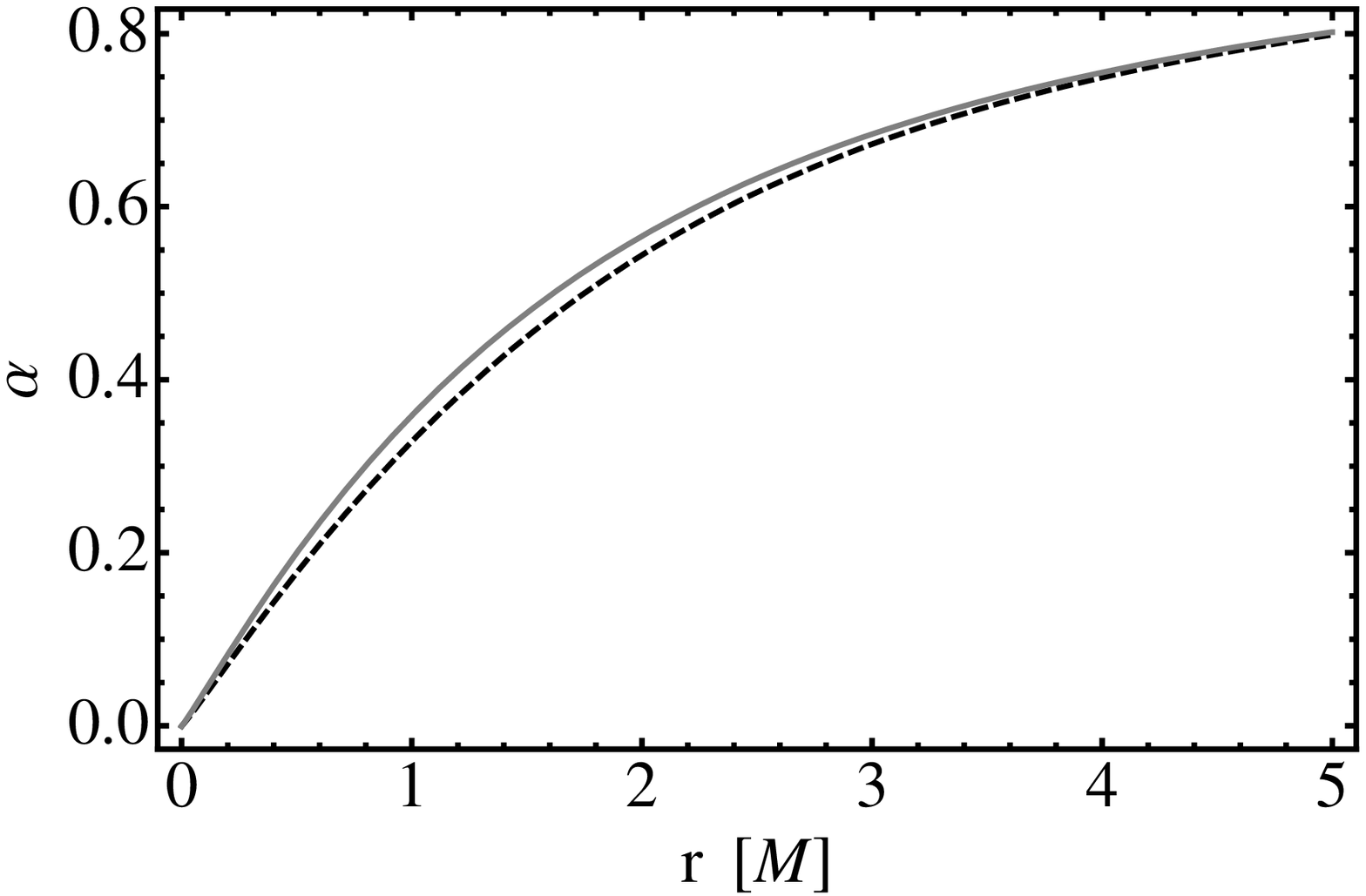}}
    \put(-100,65){\includegraphics[width=.2\textwidth]{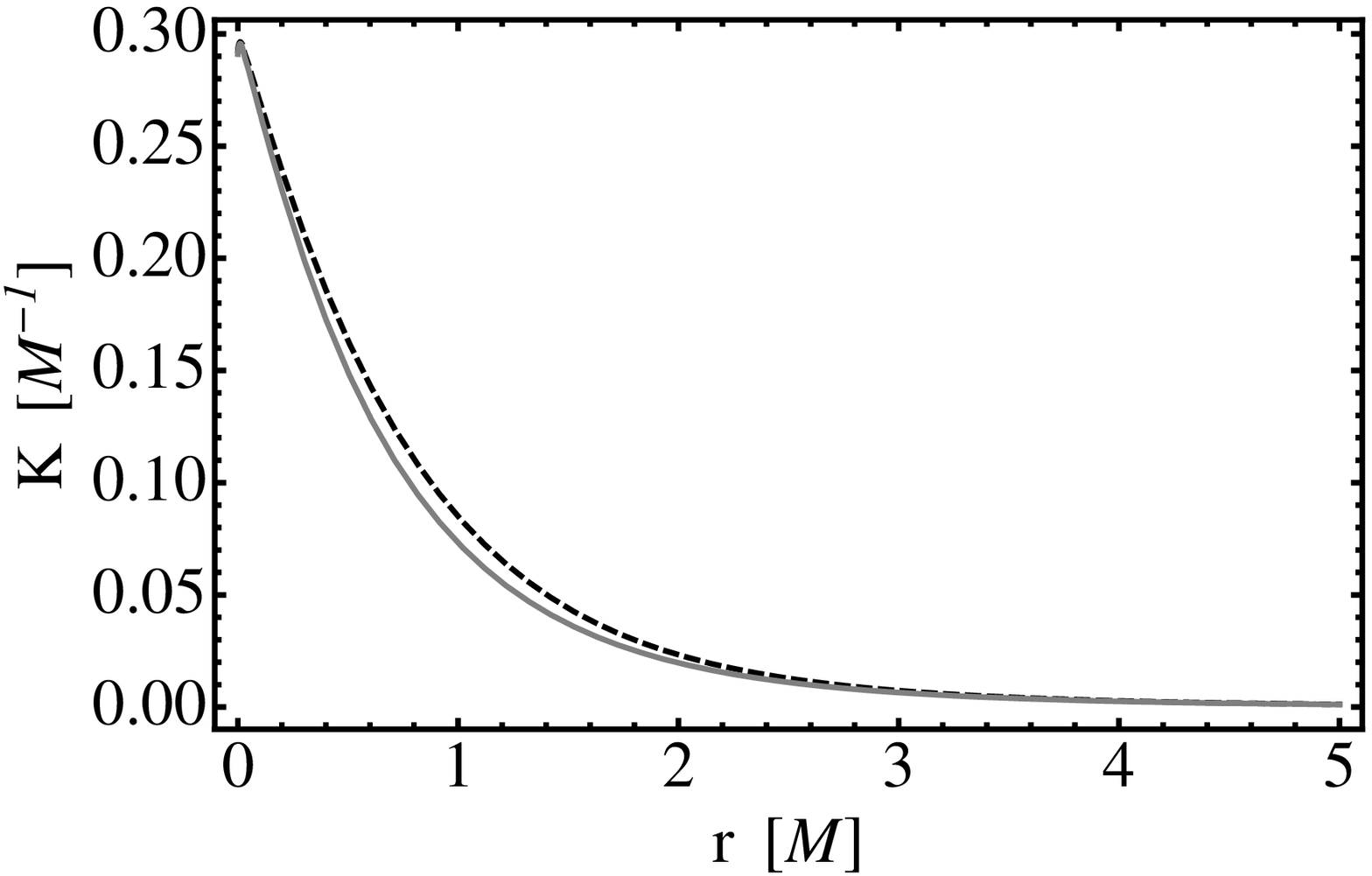}}
    \put(-175,20){\includegraphics[width=.1\textwidth]{fig2_eta_legend2}}
    \caption{The lapse function $\alpha$ as a function of extrinsic curvature
    $K$ for a single, non-spinning and non-moving puncture after a time
    $t=50\,$.M. We compare using $\er$ (black, dashed line) and $\etas=2.0/M$
    (gay line). The two curves lie  perfectly on top of each other and are
    therefore indistinguishable. The insets show  lapse (upper panel) and
    extrinsic curvature (lower panel) as functions of  distance from the
    puncture.}
    \label{fig:alphaoverK}    
  \end{figure}

  The changes in the shift should only affect the
  coordinates and coordinate independent quantities should not change.
  This can be examined by looking at a scalar as a function
  of another scalar, e.g.\ the lapse $\alpha$ as a function of extrinsic
  curvature $K$, $\alpha=\alpha(K)$. Both scalars should see the same coordinate
  drifts and therefore, no changes are expected in $\alpha(K)$. Figure
  \ref{fig:alphaoverK} confirms this expectation. The
  two curves $\alpha(K)$ for $\etas=2.0/M$ and $\er$ are lying perfectly on top
of
  each other. 
  We therefore believe that using the dynamical damping
  introduces only coordinate changes in our puncture simulations. 

  \subsection{\label{results:em}Black hole binary with equal masses}
  While Eq.~(\ref{eq:etapsimm}) has been introduced in order to allow for
  numerical simulations of two black holes with highly different masses, we
  first apply it to equal mass simulations in order to perform several
  consistency checks.
  
  The (first order) coordinate independent quantity to look at in
  binary simulations is the Newman--Penrose scalar $\Psi_4$. We use
  $\Psi_4$ for the extraction of gravitational waves (see
  \cite{BruGonHan06} for details of the wave extraction algorithm),
  decomposed into modes using spin-weighted spherical harmonics
  $Y_{lm}^{-2}$.  Since $\Psi_4$ is only first-order gauge invariant
  and we furthermore extract waves at a finite, fixed coordinate
  radius, it is a priori an open question how much the changes in the
  shift affect the wave forms.

\begin{figure}[tb]
    \centering
    \begin{overpic}[width=0.45\textwidth]{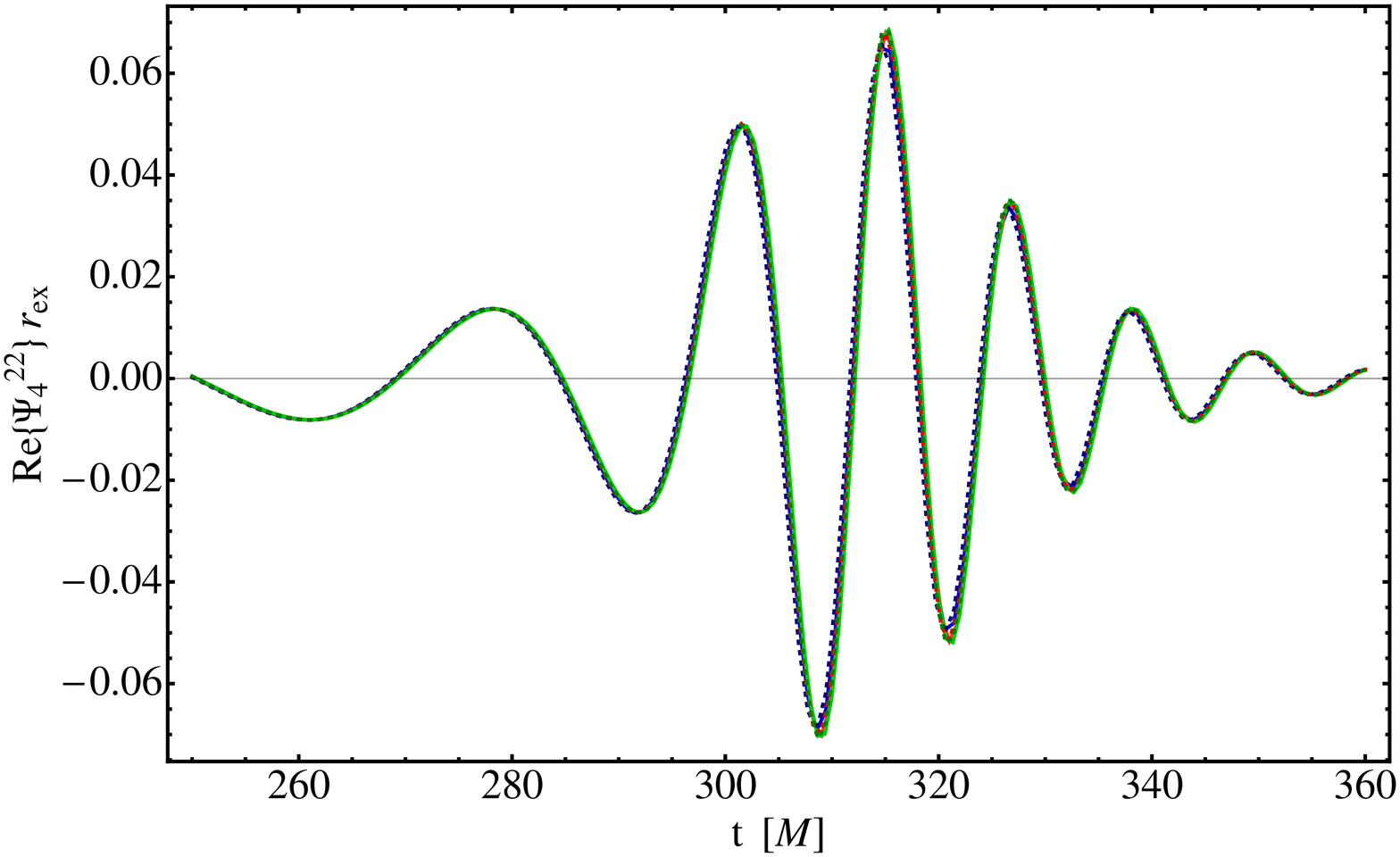}
      \put(12,40){\includegraphics[width=0.1\textwidth]{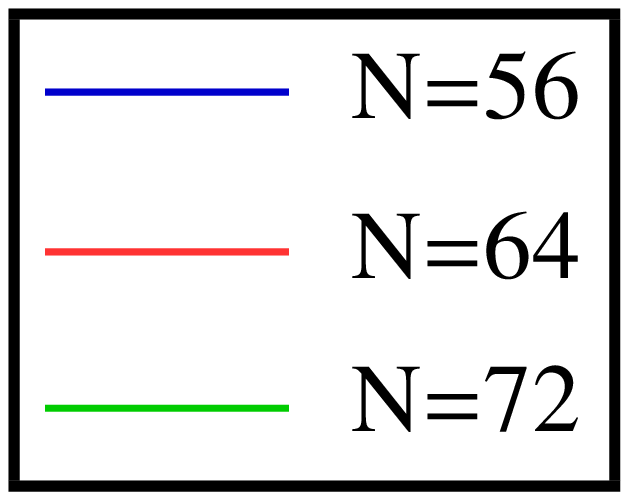}}
    \end{overpic}
    \caption{
    Real part of the 22-mode of $\Psi_4$ times extraction radius $r_{\mbox{ex}}$
    for an equal mass binary with initial separation $D=7\,M$ using
    $\etas=2.0/M$ (solid lines) and $\er$ following
    Eq.~(\ref{eq:etapsimm}) (dashed lines) in three different resolutions
    (blue,
    red, green lines) according to the grid configurations described in the
    text.}
    \label{fig:emRepsi4}
  \end{figure}

\begin{figure}[tb]
    \centering
    \begin{overpic}[width=0.45\textwidth]{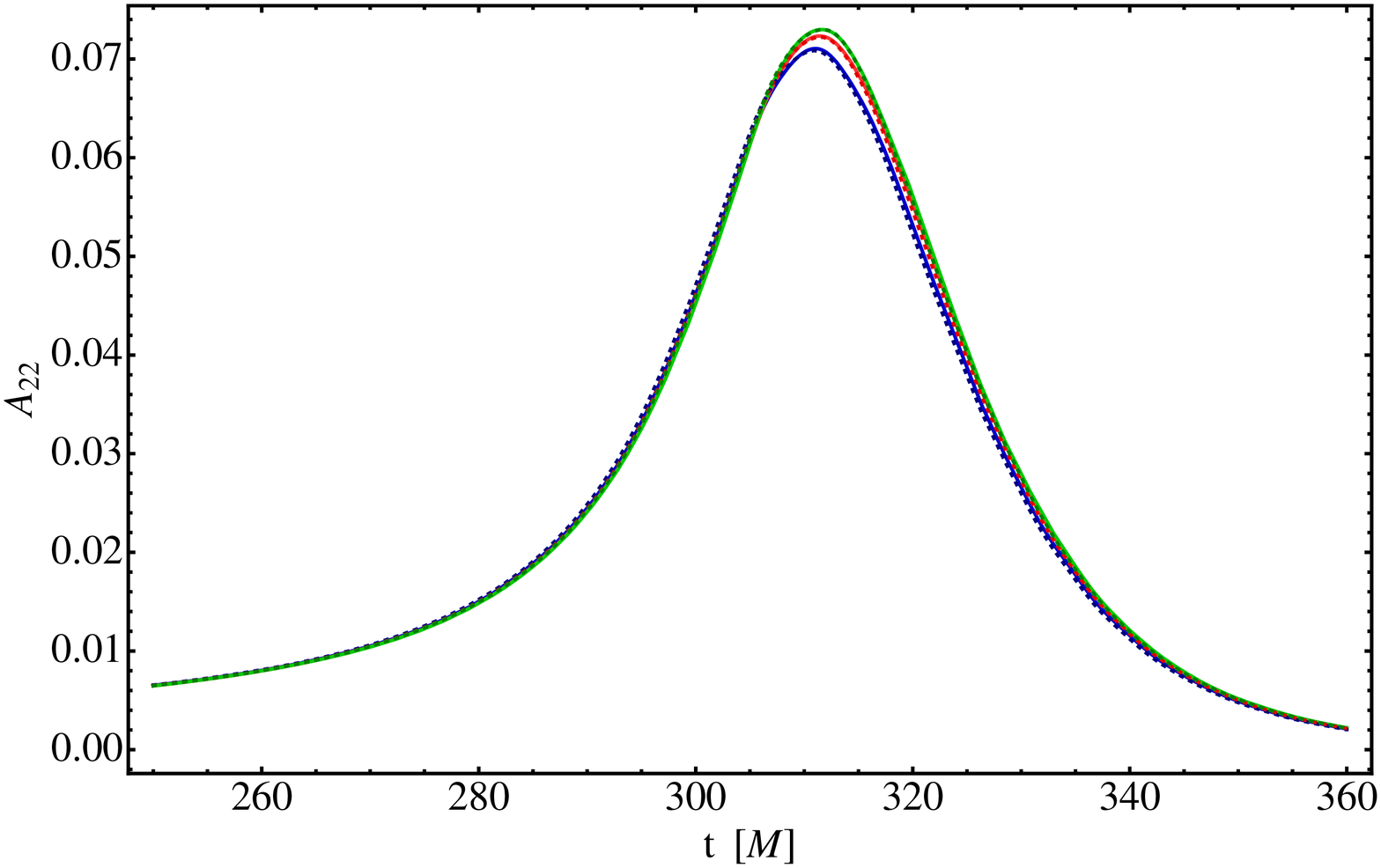}
   \put(10,9){\includegraphics[width=0.22\textwidth]{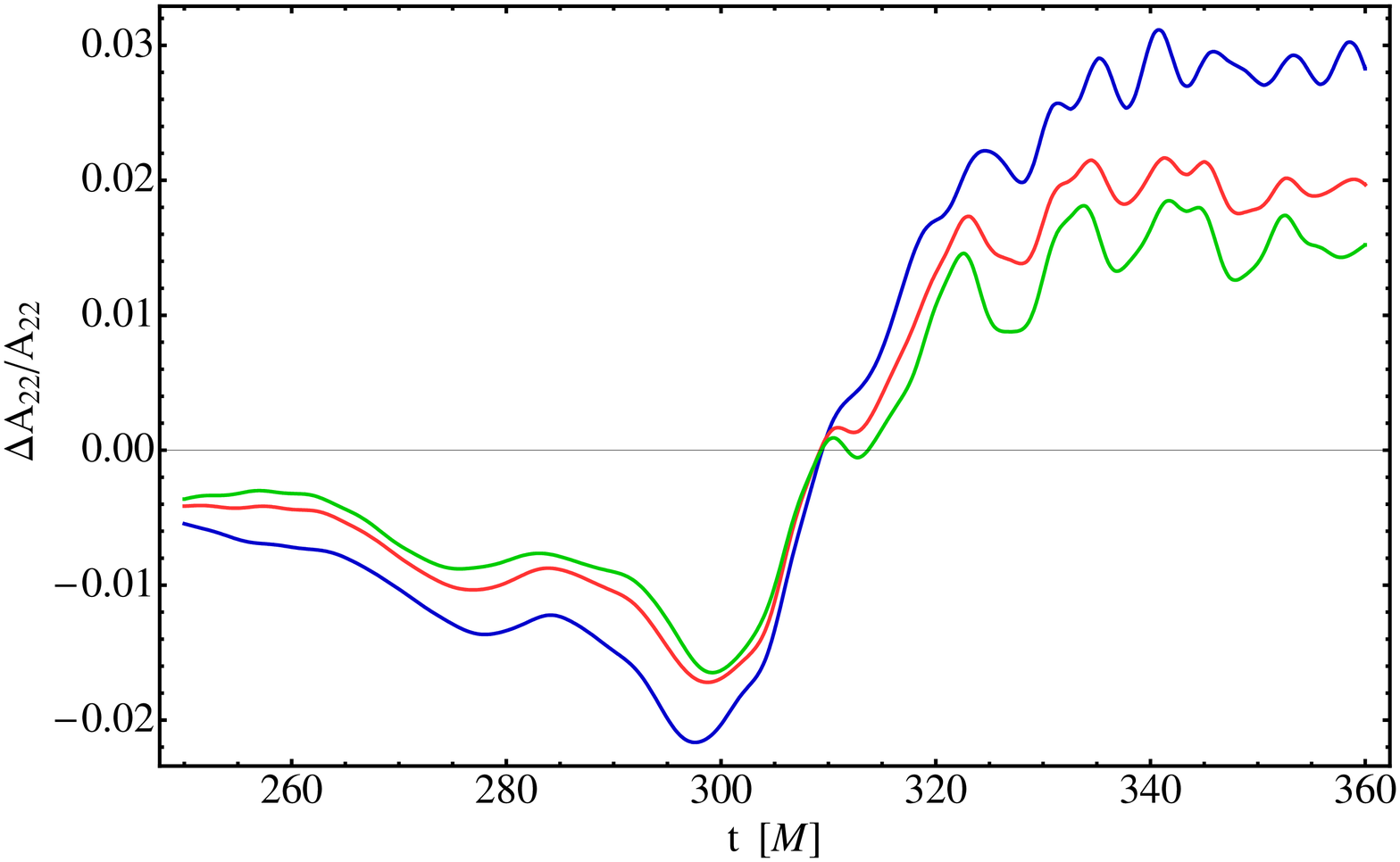}
    }
      \put(75,40){\includegraphics[width=0.1\textwidth]{fig4_conv_legend}}
    \end{overpic}
    \caption{
    Amplitude of the 22-mode of $\Psi_4$ for the same
    binary as in Fig.~(\ref{fig:emRepsi4}) using $\etas=2.0/M$ (solid lines) and
    $\er$ (dashed lines) in three different
    resolutions (blue, red, green lines) according to the grid configurations
    described in the text.
    The inset shows the relative deviation $\Delta A_{22}/A_{22}$
    between the amplitude in the standard and in the new gauge, again for the
    three different resolutions.}
    \label{fig:empsi4amp}
  \end{figure}

\begin{figure}[tb]
    \centering
    \begin{overpic}[width=0.45\textwidth]{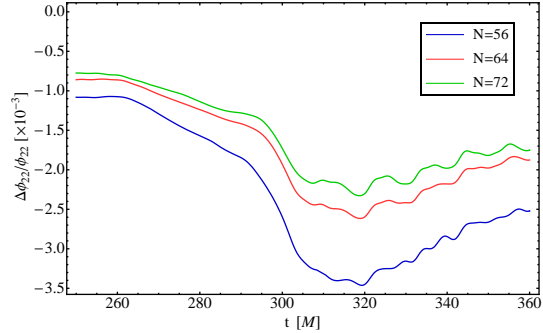}
       \put(75,40){\includegraphics[width=0.1\textwidth]{fig4_conv_legend}}
    \end{overpic}
    \caption{
    Relative phase  difference of the 22-mode of $\Psi_4$ for the same
    binary as in Fig.~(\ref{fig:emRepsi4}) using $\etas=2/M$ and $\er$. Compared
    are three different resolutions (blue, red and green lines) according to the
    grid configurations described in the text.}
    \label{fig:empsi4phase}
  \end{figure}

  As the most dominant
  mode of $\Psi_4$ in an equal mass simulation is the $l=|m|=2$ mode,  its real
  part multiplied by the extraction radius ($r_{\mbox{ex}}=90\,M$ in this case)
  is displayed in  Fig.~\ref{fig:emRepsi4}.
  We look at amplitude and phase of this mode  in Figs. \ref{fig:empsi4amp}
  and \ref{fig:empsi4phase}. The initial separation was chosen
  to be $D=7\,M$. The black holes complete about 3 orbits. 
  Three different resolutions are used corresponding to the three different
  colors in Figs.~\ref{fig:emRepsi4}, \ref{fig:empsi4amp} and
  \ref{fig:empsi4phase}. We use the number of grid points in the inner boxes
  (centered around the black holes) to denote the different resolutions.  
  The grid configurations, in the terminology of~\cite{BruGonHan06}, are
  $\phi[5\times56:5\times112:6]$, $\phi[5\times64:5\times128:6]$, and
  $\phi[5\times72:5\times144:6]$ which corresponds to resolutions on the finest
  grids of $3\,M/112$, $3\,M/128$ and $M/48$, respectively. These are the grid
  configurations used in~\cite{HusGonHan07}.
 
  In Fig.~\ref{fig:empsi4amp}, we compare the amplitude $A_{22}$ in the standard
  gauge, $\etas=2.0/M$, displayed as  solid lines, to the new one,
  Eq.~(\ref{eq:etapsimm}), plotted as dashed  lines. 
  We find that the differences between standard and new gauge for a given grid
  resolution are much smaller than differences due to using different
  resolutions. 
  This strengthens the belief that we only introduced coordinate
  changes to the system when using Eq.~(\ref{eq:etapsimm}). The maximum relative
  deviation between the amplitudes
  $A_{22}$ of old and new gauge amounts to about $3\%$ for the
  lowest resolution ($N=56$) and decreases with increasing resolution, which can
  be seen in the inset of Fig.~\ref{fig:empsi4amp}. 
  For the phase, the absolute differences are not visible by eye and therefore,
  we only plot the relative deviations between $\phi_{22}$ in the standard and
  new gauge for the three resolutions. For the lowest resolution ($N=56$), the
  maximum deviation is only $0.35\%$. 
  As for the amplitude, this deviation decreases with increases in resolution.
  This shows how the differences in the waveforms disappear with increasing
  resolution. 

  The fact that there actually {\em are} differences visible in the
  waves, though very small ones, is not surprising when considering
  the way we extract gravitational waves.  We fix a certain extraction
  radius and compute the Newman--Penrose scalar on a sphere of this
  radius.  The radius itself is coordinate dependent and we are
  comparing $\Psi_4$ extracted at slightly different radii in the
  standard and new gauges. In future work we plan to compare wave forms
  extrapolated in radius to infinity, although it is worth noting how
  small the deviations are without additional processing.


  \subsection{\label{results:um4}Black hole binary with mass ratio 4:1}
  
  After having examined the influence of using a dynamical damping coefficient
  $\er$ for an equal mass binary, the next step is to look at its behavior for
  unequal masses. The following results are obtained from a simulation of two
  black holes with mass ratio $q=4$ and initial separation $D=7\,M$.
  We used the grid configurations $\phi[5\times N:7\times 2N:6]$ with $N=72,
  80$ which have also been used in \cite{DamNagHan08} for mass ratio 4:1.
  An interesting question in this context is how Eq.~(\ref{eq:etapsimm})
  behaves for a simulation with two punctures with different masses. 
  The analytical behavior (\ref{eq:eta_limits}) was deduced for a single,
  non-moving, stationary puncture but now we are using it for two moving
  punctures, which are during most of the simulations far from having reached a
  stationary state globally, but approximately stationary locally at
  the punctures. 

\begin{figure}[tb]
    \centering
    \includegraphics[width=0.45\textwidth]{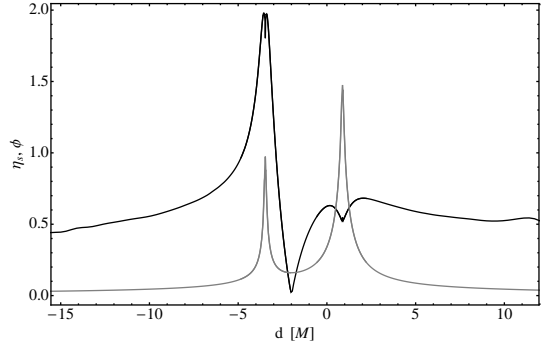}
    \caption{
    $\er$ (black line) for two  punctures of bare masses $M_1=0.5$ and
    $M_2=2.0$   and the conformal factor $\phi$    (gray line) whose
    maxima show the current positions of the punctures. 
In this plot the $M_i$ are dimensionless, so $\er$ is dimensionless as well.
The snapshot is taken
    right after the beginning of the $q=4$ simulation with initial separation
    $D=5\,M$ described in the text. The smaller black hole is at position
    $d=-3.6\,M$, the larger one at position $d=1.2\,M$. 
}
    \label{fig:etaoverpos}
  \end{figure}

  Figure~\ref{fig:etaoverpos} illustrates the distribution of $\er$
  between the two punctures. For convenience, the conformal factor
  $\phi = \ln \psi$ is also plotted in order to indicate the positions
  of the punctures via its maxima (the divergences are not resolved).
  The snapshot is taken at a time during the simulation when the
  punctures are still well separated. Similar to the simulations of a
  single puncture, according to (\ref{eq:eta_limits}) we expect to
  find $\er\simeq 2$ near the puncture with mass $M_1=0.5$ and
  $\er\simeq 0.5$ in the vicinity of the second puncture with
  $M_2=2.0$.  Near the outer boundary, $\etas$ is supposed to take the
  value $1.31/(M_1 + M_2) = 0.52$. Here the $M_i$ are chosen to be
  dimensionless, so $\er$ is dimensionless as well.
  Figure~\ref{fig:etaoverpos} confirms that we do obtain the expected
  values, although they are not reached exactly. The latter is not a
  problem as simulations work nicely as long as $\er$ is in the right
  range for each black hole.  For this reason, Eq.~(\ref{eq:etapsimm})
  also seems to work rather nicely for two punctures with unequal
  masses.
  
\begin{figure}[tb]
    \centering
    \begin{overpic}[width=0.45\textwidth]{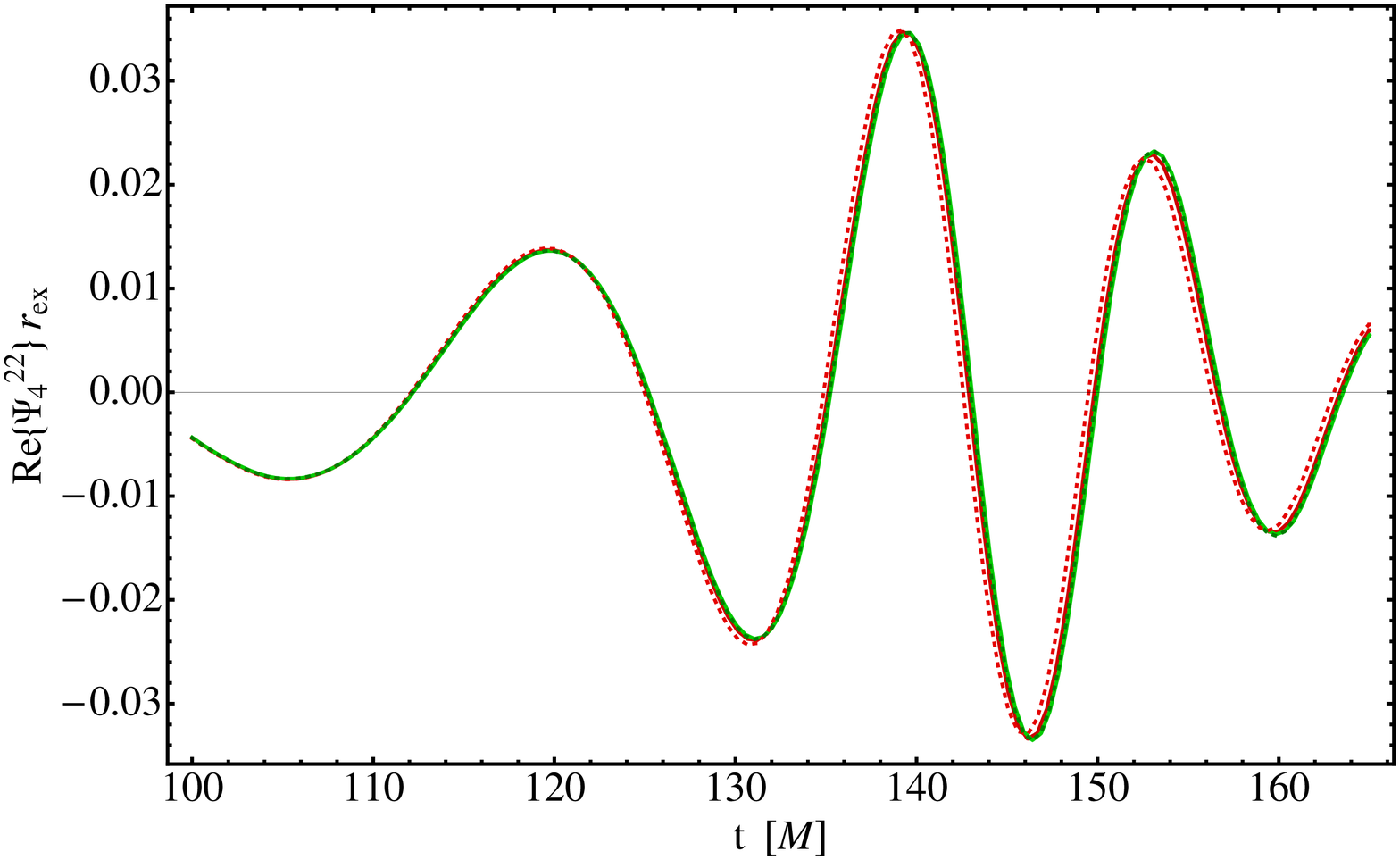}
    \put(13,42){\includegraphics[width=0.1\textwidth]{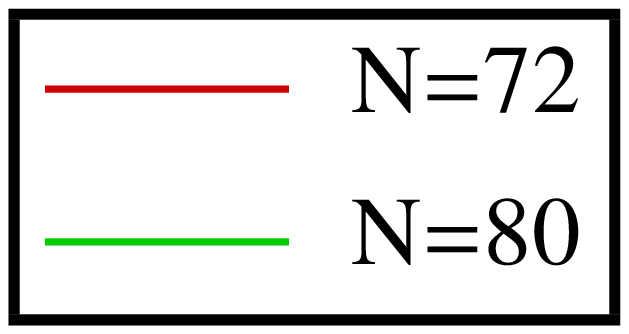}}
    \end{overpic}
     \caption{
     Real part of the 22-mode of $\Psi_4$ multiplied by the extraction radius
     $r_{\mbox{ex}}$ for $q=4$ and initial separation $D=5\,M$ runs.
    Compared are results for employing $\etas=2.0/M$ (solid lines) and $\er$
    (dashed lines) in two different resolutions (red and green lines)
    according
    to the grid configurations described in the text.
      }
    \label{fig:um4Repsi4}
  \end{figure}

\begin{figure}[tb]
    \centering
    \begin{overpic}[width=0.45\textwidth]{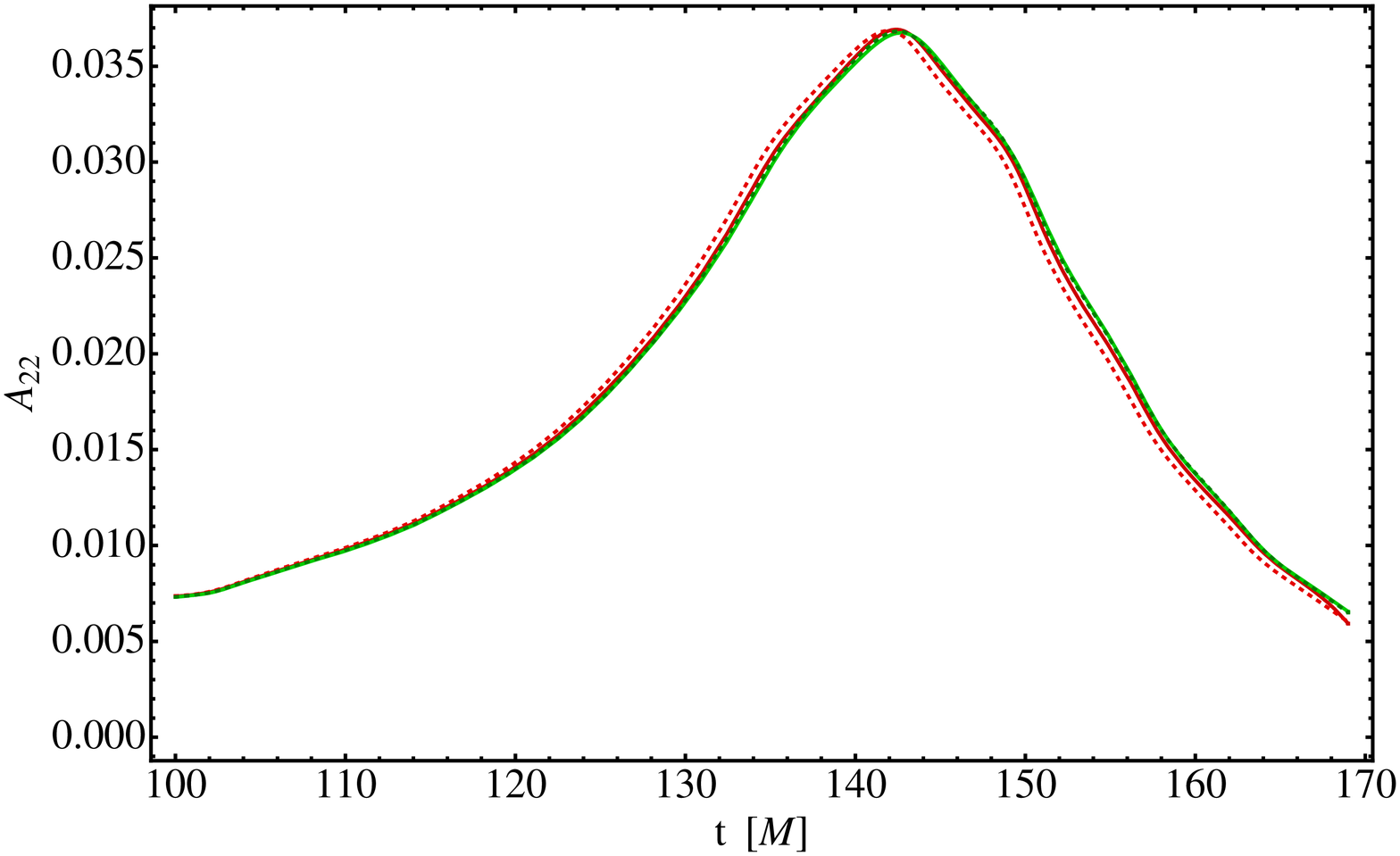}
    \put(12,9){\includegraphics[width=0.22\textwidth]{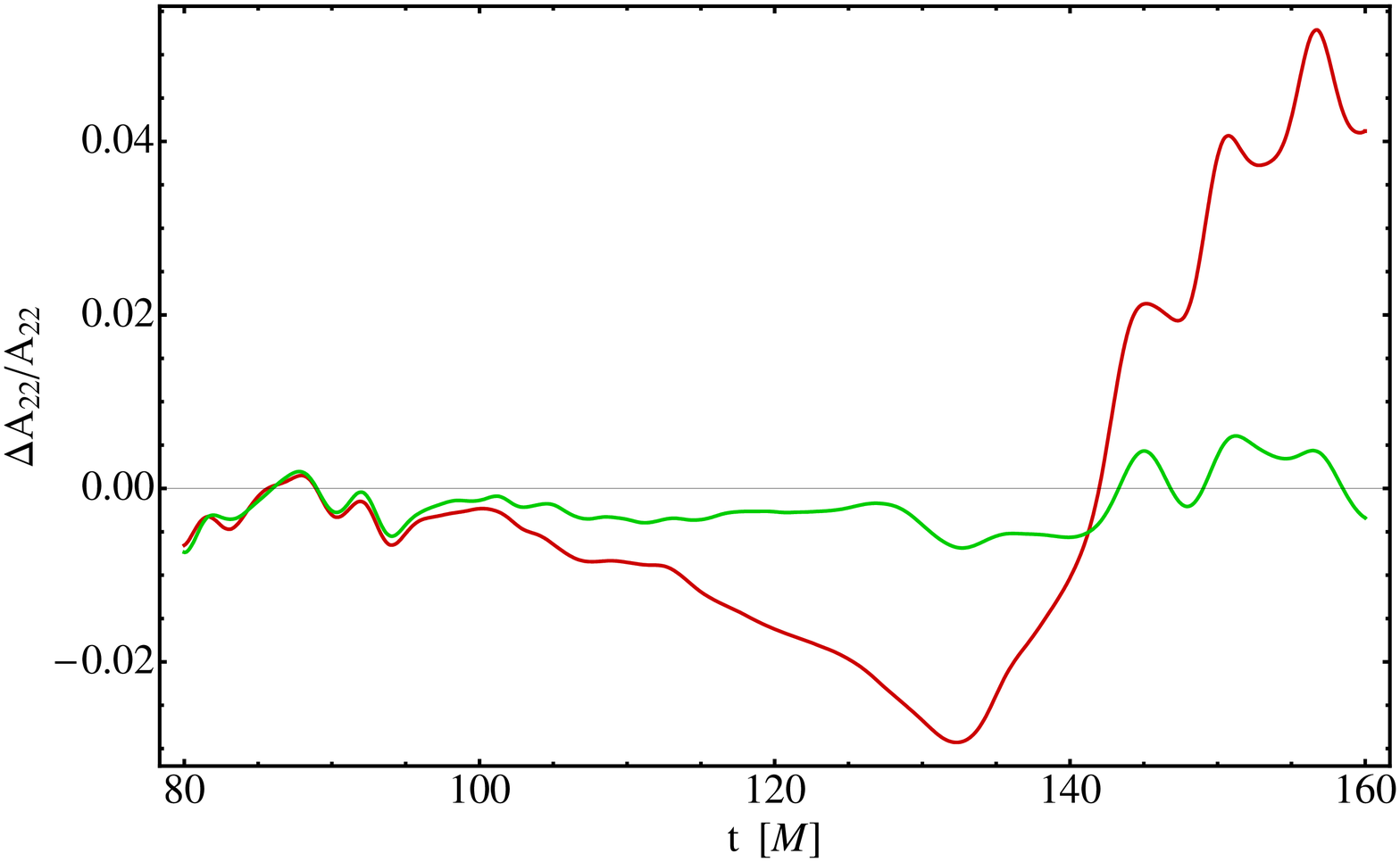}
    }
    \put(75,42){\includegraphics[width=0.1\textwidth]{fig7_conv_legend_2res}}
    \end{overpic}
    \caption{
    Amplitude of the 22-mode of $\Psi_4$ for the same runs as in
    Fig.~\ref{fig:um4Repsi4}.
    We compare results using $\etas=2.0/M$ (solid lines) and $\er$
    (dashed lines) in two different resolutions (red and green lines) according
    to the grid configurations described in the text. The inset shows the
    relative deviation $\Delta A_{22}/A_{22}$ between the amplitude in the
    standard and in the new gauge, again for the same two  resolutions.}
    \label{fig:um4psi4amp}
  \end{figure}

\begin{figure}[tb]
    \centering
    \begin{overpic}[width=0.45\textwidth]{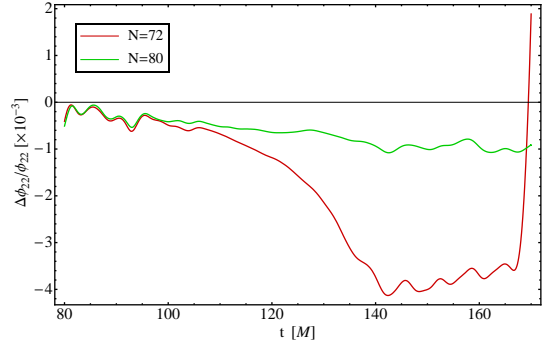}
       \put(10,44){\includegraphics[width=0.1\textwidth]{fig7_conv_legend_2res}}
    \end{overpic}
    \caption{Relative phase difference of the 22-mode of $\Psi_4$ for the same
    run as in Fig.~\ref{fig:um4Repsi4} $\etas=2.0/M$ and $\er$.
    Compared are two different resolutions according to the grid
    configurations described in the text.}
    \label{fig:um4psi4phase}
  \end{figure}

  As we did in the equal mass case in section \ref{results:em}, we compare the
  22-mode of $\Psi_4$ in the new gauge with the standard gauge.
  To see how small the differences using $\etas$ or $\er$ are, we plot its real
  part using two different resolutions which correspond to the two different
  colors in Fig.~\ref{fig:um4Repsi4}.
  Figure~\ref{fig:um4psi4amp} shows the amplitude for the two different
  resolutions. The inset gives the relative differences between amplitudes in
  the standard and new gauge. The maximum relative deviation appears for the
  lower resolution and amounts to about $3\%$. The high resolution gives  
  $0.5\%$ relative difference. The phases in standard and new  gauge are
  compared in Fig.~\ref{fig:um4psi4phase}. Again, we show only the  relative
  deviations as the absolute ones are too small to be seen. We  find relative
  differences of up to $0.4\%$ for the lower resolution and only  $0.1\%$ for
  the high one. This confirms that we are changing only the  coordinates, as we
  found before in Sections \ref{results:schwarzschild} and  \ref{results:em}.

  
\begin{figure}[t]
  \centering
  \includegraphics[width=0.45\textwidth]{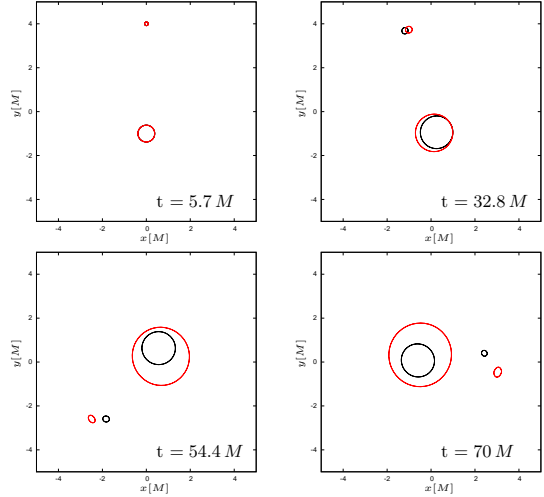}
  \caption{
  Comparison of apparent horizons in the orbital plane using $\etas=2.0/M$ (red
  lines) and $\er$ (black lines) at different times during the evolution for a
  $q=4$ run with initial separation $D=5\,M$.}
  \label{fig:um4AH}
  \end{figure}

\begin{figure}[t]
  \centering
  \includegraphics[width=0.45\textwidth]{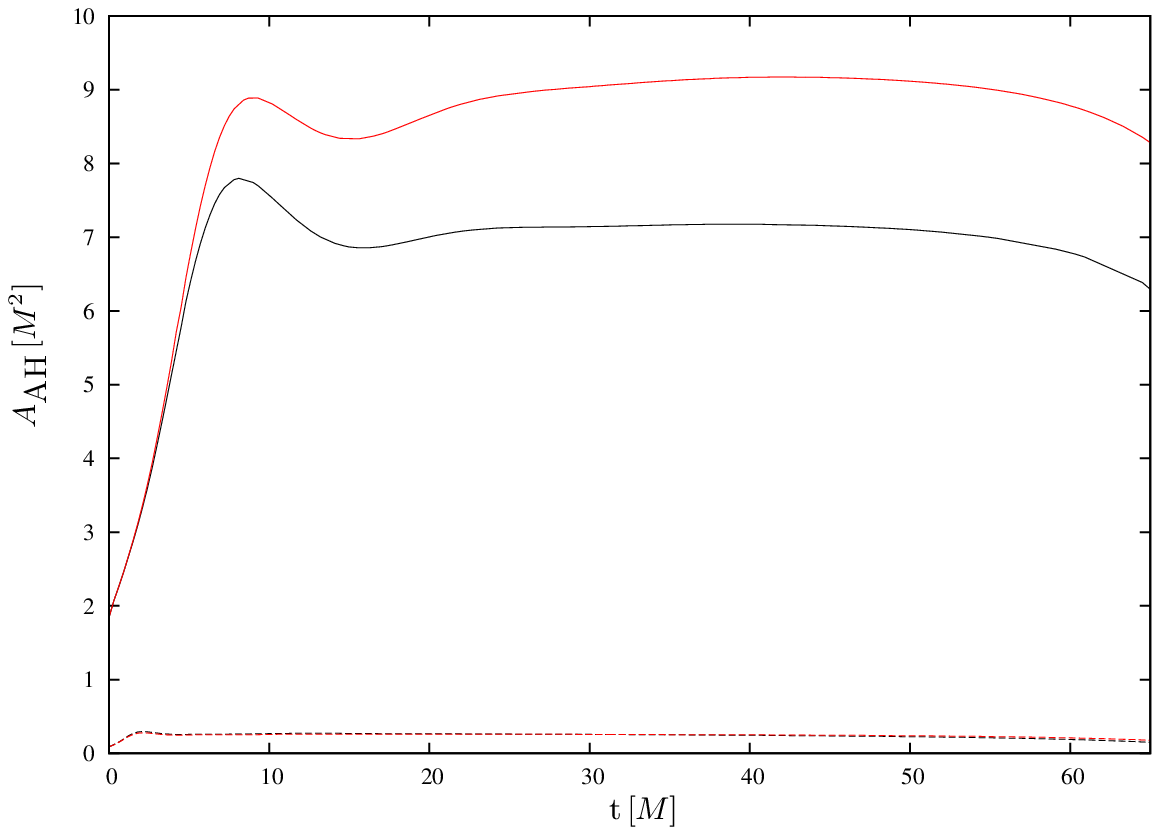}
  \caption{
  Comparison of the coordinate area of the apparent horizons using $\etas=2.0/M$
  (red lines) and $\er$ (black lines) over evolution time for a
  $q=4$ run with initial separation $D=5\,M$ until shortly before a common
  apparent horizon appears. The dashed lines belong to the smaller black hole
  whereas the solid lines represent the larger black hole.}
  \label{fig:um4AHcA}
  \end{figure}

While the invariance of the waveforms is the most important feature of
the new gauge $\er$, it is illuminating to examine how the black holes
are represented on the numerical grid. To this end, the apparent
horizons (AH) are computed for both gauges in Fig.~\ref{fig:um4AH}.
We show the result in the $(x,y)$--plane, in which the
orbital plane lies. For clarity, the slices through the apparent horizons are
only shown at 4 different times. In the beginning of the simulations, the AH
pertaining to the same black hole are lying on top of each other. With time,
they separate as the coordinates become more and more different in the two 
simulations. Two observations can be made.
First, the ratio between the coordinate area of the AH of the larger black hole
and the one of the smaller black hole is larger in the simulation using
$\etas=2.0/M$. This means the black holes are represented more equally on the
grid
in the simulation using Eq.~(\ref{eq:etapsimm}). This fact can be seen even
more clearly in Fig. \ref{fig:um4AHcA} where we plot the coordinate area of the
apparent horizons comparing the standard gauge (red lines) and the new one
(black lines). While the coordinate sizes of the smaller black hole (dashed
lines) are nearly equal in both gauges, the sizes of the larger black hole
(solid lines) differ by roughly $2\,M^2$. 
Second, the shape of the horizon of the smaller black hole is more and more
distorted in the $\etas=2.0/M$-simulation when the black holes come closer
together. This deformation is not visible in the new coordinates. The
progressive stretching of the apparent horizon shape and therefore the
distortion
of the coordinates near the black holes can be a source
of instabilities, e.g.\ \cite{BruTicJan03}. Using
Eq.~(\ref{eq:etapsimm}) seems to be profitable in this regard.

  \subsection{\label{sec:problem}Behavior of $\er$ and influence on the shift
  vector}
\begin{figure}[t]
    \centering
    \includegraphics[width=0.45\textwidth]{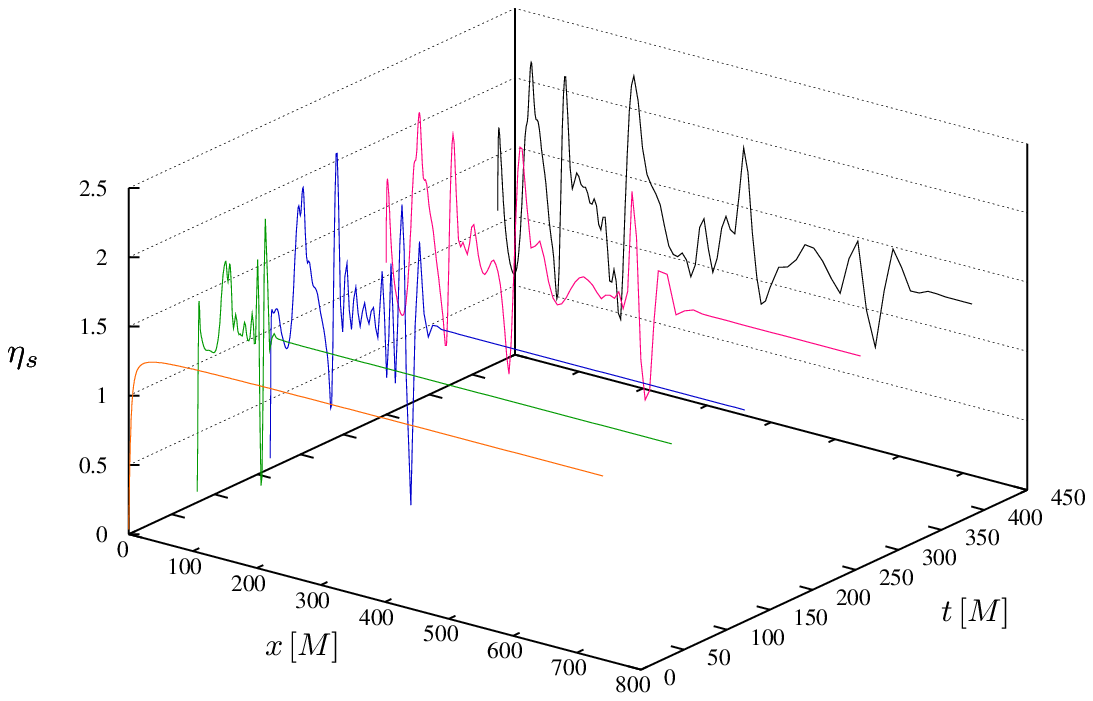}
    \caption{
    Form of $\er$ in $x$-direction at different times during an equal
    mass binary simulation. Noise travels outwards and leaves strong
    distortions on the grid.}
    \label{fig:emcompEtaT}
  \end{figure}

\begin{figure}[t]
    \centering
    \includegraphics[width=0.45\textwidth]{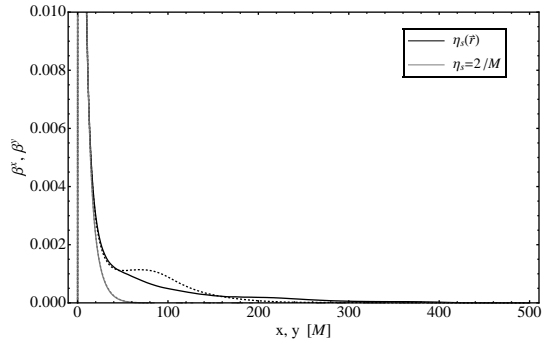}
    \put(-55,95){\includegraphics[width=.1\textwidth]{fig2_eta_legend2}}
    \caption{
    $x$-component of the shift vector in $x$-direction (solid curves)
    and $y$-component of the shift vector in $y$-direction (dashed curves)
    after the merger of two equal mass black holes at time $t=500\,M$ when using
    either the standard gauge $\etas=2.0/M$ (gray lines) or the dynamical one,
    $\er$, (black lines).} 
    \label{fig:em_betax_r}
  \end{figure}

  Despite the encouraging results we have seen so far, there is a
  non--negligible
  concern using Eq.~(\ref{eq:etapsimm}) in the gamma-driver condition
  (\ref{eq:gammadriver}). Although we do not determine the damping coefficient
  via a wave equation, we see wavy features in $\er$ traveling outwards.
  These distortions even leave remnants on the grid, especially when they pass
  through a refinement boundary. The form of $\etas (x)$ after different
evolution
  times can  be seen in Fig. \ref{fig:emcompEtaT} for the equal mass binary
  described in  Sec.  \ref{results:em}. The result is similar in the $q=4$
  simulation and even  in the  Schwarzschild simulation, an outward traveling
  pulse is present,  which  however  does not leave visible distortions on the
  grid and the relative  amplitude of  which decreases for higher mass. 
  The effort we made before in order to achieve the correct value of $\er$ near
  the outer boundary seem to be canceled out by the disturbed shape we find
  now. 
  As the peaks travel to a region of the grid where we have no punctures, we
  might take the point of view that the exact value of $\er$ and therefore the
  distortions are of no importance for our simulations. Indeed the oscillations
  do not translate to oscillations in the shift vector as one might think. In
  the shift, we find  no gauge ``waves'' related to the ones in $\er$.
  Nevertheless, there is an unusual behavior. After merger,
  when going away from the punctures the shift does not fall off to zero as fast
  as it does when using $\etas=\mbox{const.}$ but keeps a shoulder (compare
  Fig.~\ref{fig:em_betax_r}) which might lead to an unphysical and unwanted
  drift of the coordinate system.
  We are planning to investigate these issues in more detail in the future.


\section{\label{discussion}Discussion}

We presented a new approach to determine the coordinates in slices
of spacetime for binary black hole simulations where we take the
distribution of mass over the grid into account. We have shown that
our approach of determining the damping parameter in the gamma-driver
condition dynamically via Eq.~(\ref{eq:etapsimm}) gives stable
evolutions and does not significantly change the gravitational waves
extracted from binary systems of equal or unequal masses. Furthermore,
the use of Eq.~(\ref{eq:etapsimm}) in an unequal mass simulation
resulted in a more regular shape of the apparent horizon of the
smaller black hole as the binary merges. The coordinate size of the
apparent horizons became more uniform with the new damping coefficient
which is a first step towards representing and resolving black holes
with different masses equally and hence removing the large asymmetry
which usually distorts the numerical grid in unequal mass simulations.
We found gauge waves in our damping coefficient which might affect the
stability in very long-term simulations and lead to coordinate drifts
after the merger of the binary. We will address these issues in a
future publication \cite{MueGriBru09}.


\ack
It is a pleasure to thank Jason Grigsby for discussions and for his valuable
comments on this publication.
We also thanks David Hilditch for discussions on the hyperbolicity of the BSSN
system. 
This work was supported in part by DFG grant SFB/Transregio~7
``Gravitational Wave Astronomy'' and the DLR (Deutsches Zentrum f\"ur Luft
und Raumfahrt).  Doreen M\"uller was additionally supported by the  DFG Research
Training Group 1523 ``Quantum and Gravitational Fields''.
Computations were performed on the HLRB2 at LRZ Munich.

\section*{References}

\end{document}